\pdfoutput=1

\documentclass[aps,prl,twocolumn,showpacs,amsfonts,amsmath,floatfix]{revtex4-1}

\usepackage{graphicx}

\begin{document}


\title{Tunneling theory of two interacting atoms in a trap}

\author{Massimo Rontani}
\email{massimo.rontani@nano.cnr.it}
\affiliation{CNR-NANO Research Center S3, Via Campi 213a, 41125 Modena, Italy}




\date{\today}

\begin{abstract}
A theory for the tunneling of one atom out of a trap
containing two interacting cold atoms is developed.
The quasiparticle wave function, dressed by the interaction
with the companion atom in the trap, replaces the non-interacting
orbital at resonance in the tunneling matrix element.
The computed decay time for two $^6$Li atoms
agrees with recent experimental results  
[G. Z\"urn, F. Serwane, T. Lompe, A. N. Wenz,
M. G. Ries, J. E. Bohn, and S. Jochim, arXiv:1111.2727],
unveiling the `fermionization' of the wave function 
and a novel two-body effect.
\end{abstract}


\maketitle
Recently few cold atoms have been confined
in tight single optical traps with control of their number
and quantum states \cite{Cheinet08, Serwane11}. 
This capability is exciting as it 
bridges different fields such as mesoscopic and nuclear physics.
Systems made of just two interacting atoms are especially relevant, 
being the building blocks of many-body strongly correlated states 
\cite{BlochRMP}
and allowing for comparison with fully understood theoretical models
\cite{Busch, Girardeau10}. 

\begin{figure}
\begin{center}
\includegraphics[trim=0 18 0 0,width=8.1cm]{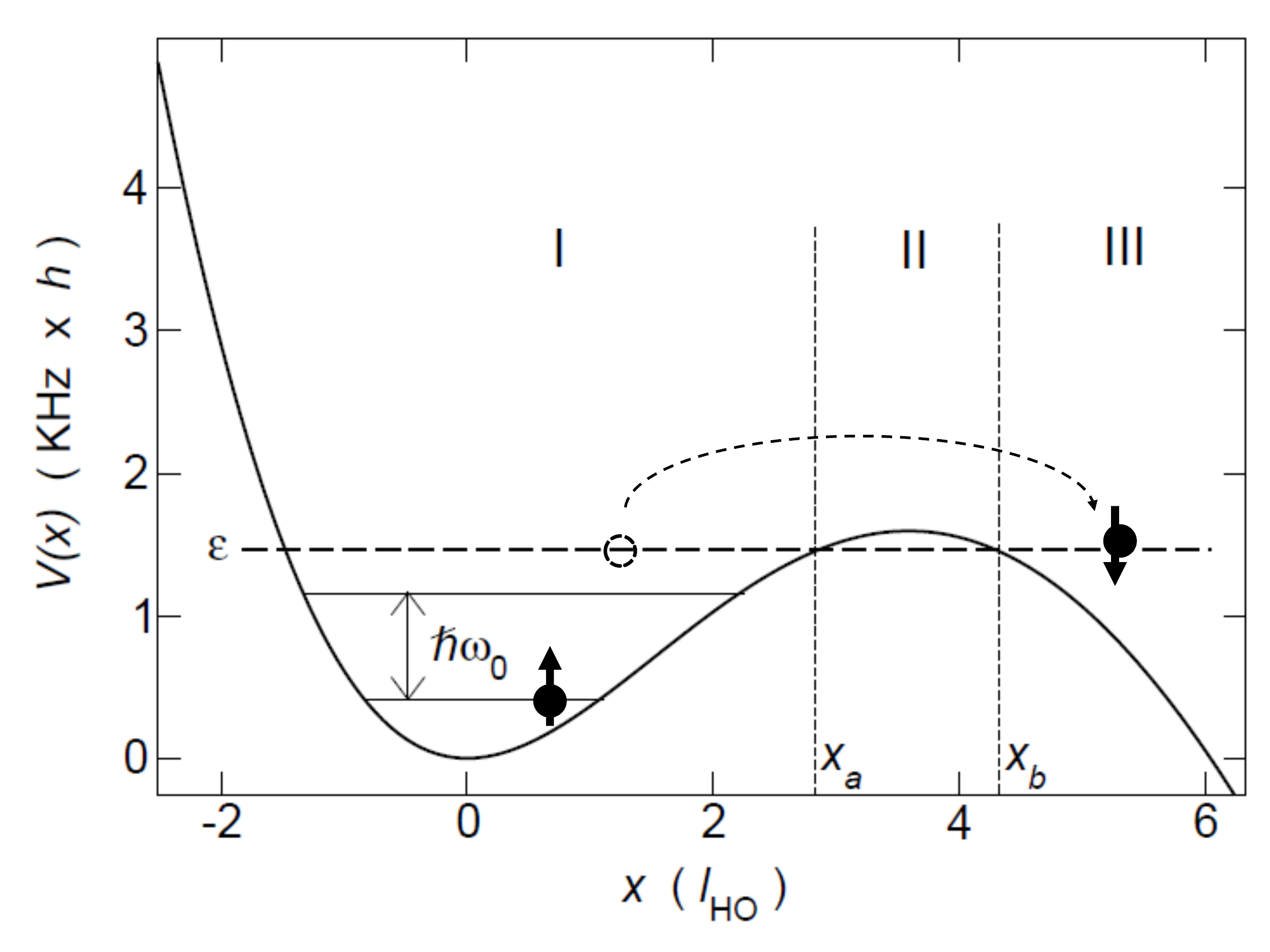}
\end{center}
\caption{ 
Confinement potential $V(x)$ vs $x$ according to Ref.~\onlinecite{Zuern11}.  
The harmonic oscillator frequency $\omega_0=$ 0.744 KHz$\cdot 2\pi$
is obtained from the spacing between the two bound states of $V(x)$
computed through the WKB approximation
(solid thin lines); $\varepsilon$ is the resonating tunneling energy. 
The length unit is $\ell_{\text{HO}}=(\hbar/m\omega_0)^{1/2}$.
The circles with arrows schematize the atoms in the
$\left|\uparrow\downarrow\right>$ configuration.
\label{frame}}
\end{figure}

\begin{figure}
\begin{center}
\includegraphics[trim = 0 60 0 60,width=8.1cm]{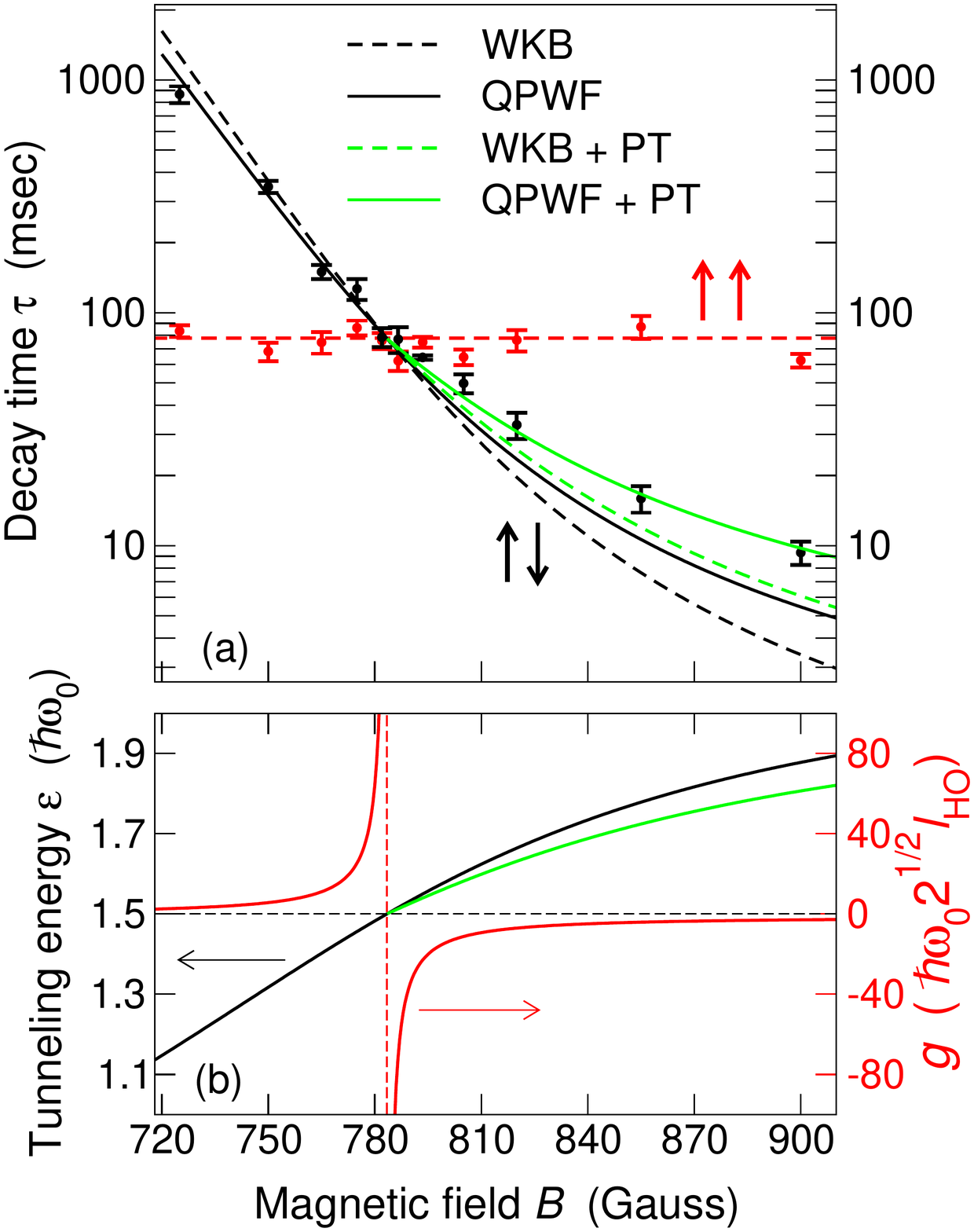}
\end{center}
\caption{(Color online) (a) Decay time $\tau$ vs magnetic field $B$.
The points with error bars are the experimental data  
\cite{Zuern11}, the dashed and solid lines are respectively 
the WKB ($\tau_0$) and QPWF predictions. 
The green [light gray] lines include the PT correction 
to the tunneling energy $\varepsilon$.
(b) Interaction strength $g$
(red [gray] curve) and $\varepsilon$ (black and green [light gray]
curve) vs $B$. 
$g$ is taken from Ref.~\onlinecite{Zuern11} and
$\varepsilon$ is computed after Ref.~\onlinecite{Busch}  
(black curve). The green (light gray) line includes the
PT correction.
The dashed lines are guides to the eye.
\label{tau}}
\end{figure}

In a fascinating experiment \cite{Zuern11},
the Heidelberg group has prepared in a one-dimensional trap 
(Fig.~\ref{frame})
two $^6$Li atoms behaving as fermions of spin one-half and
interacting through a tunable contact potential,
$g\,\delta(x_1-x_2)$.
The $\left|\uparrow\uparrow\right>$
($\left|\downarrow\downarrow\right>$) state has
a fermionic-like orbital wave function,
being odd under particle exchange, $\psi_{\uparrow\uparrow}(x_1,x_2) =
- \psi_{\uparrow\uparrow}(x_2,x_1)$, whereas the  
$\left|\uparrow\downarrow\right>$ wave function is bosonic-like,
$\psi_{\uparrow\downarrow}(x_1,x_2) =
\psi_{\uparrow\downarrow}(x_2,x_1)$.
The time $\tau$ spent by
one atom to tunnel out of the trap is measured 
as a function of the interaction strength $g$. 
As shown in Fig.~\ref{tau}, at $g=\infty$
the decay time
$\tau$ is the same for both $\left|\uparrow\downarrow\right>$ and
$\left|\uparrow\uparrow\right>$  
($\left|\uparrow\uparrow\right>$, being non-interacting,  
is independent from $g$).
The coincidence of $\tau$ is attributed to the Fermi-Bose 
duality (FBD, aka `fermionization') \cite{Girardeau60,duality,delCampo},
i.e.,
the energy and wave function modulus of $\left|\uparrow\downarrow\right>$
are the same as $\left|\uparrow\uparrow\right>$, 
$\left|\psi_{\uparrow\downarrow}(x_1,x_2)\right|^2=
\left|\psi_{\uparrow\uparrow}(x_1,x_2)\right|^2$.
This duality was predicted long ago for the Tonks-Girardeau (TG) gas 
\cite{Girardeau60} and it has been demonstrated  
for repulsively interacting bosons \cite{repbos}.
At finite $g$ both TG ($g>0$) and 
super-TG ($g<0$) states \cite{Girardeau10,Tempfli09} of 
$\left|\uparrow\downarrow\right>$ are accessed,
the former being the ground state and 
the latter the first excited state,  
here stabilized by the absence of three-body collisions \cite{Zuern11}.

The experiment raises two 
issues: (i) A rigorous argument for the coincidence 
of $\tau$ was not given.
In fact, the FBD applies only to those observables 
that involve
the wave function square modulus \cite{Girardeau60},  
whereas $\tau$ originates from the convolution of orbital 
states inside and outside 
the trap \cite{Bardeen61}. Indeed, in two dimensions $\tau$ 
may depend on the wave function phase \cite{Lorke10, Rontani11}.  
(ii) So far $\tau$ has been computed using a single-particle
theory \cite{Zuern11}, including the effect of interactions 
through energy renormalization [black dashed lines in Fig.~\ref{tau}(a)].
As the agreement with the experimental data (points) 
is qualitative, question arises whether
the shape of the interacting wave function may affect $\tau$. 
 
\begin{figure}
\begin{center}
\includegraphics[trim = 0 60 0 50, width=9.1cm]{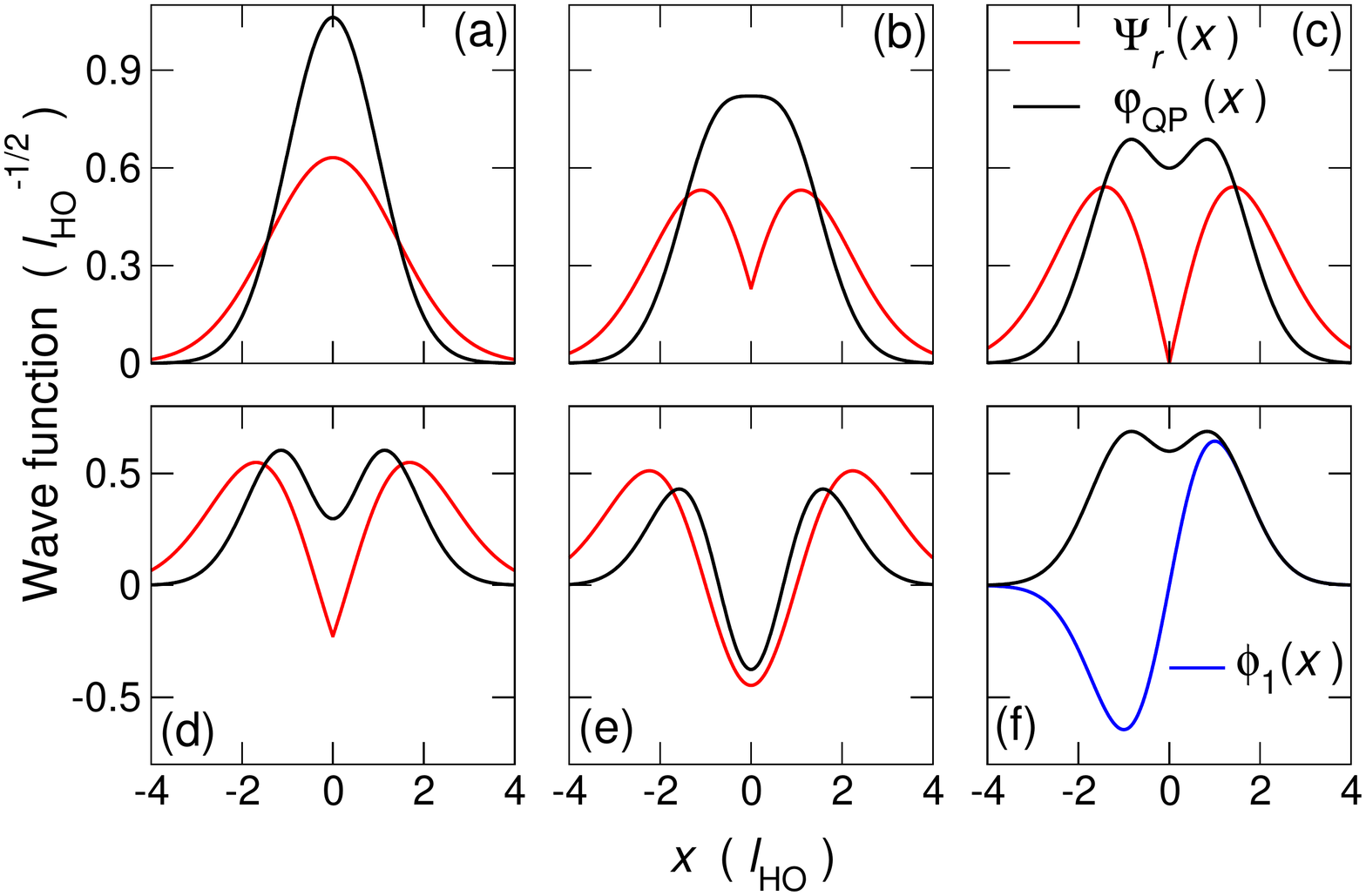}
\end{center}
\caption{(Color online)
Quasiparticle wave function $\varphi_{\text{QP}}(x)$ (black curves)
and relative-motion wave function $\Psi_r(x)$ (red [gray] curves)
of $\left|
\uparrow\downarrow\right>$ at various $g$.
(a) $g=0+$. (b) $g/(2^{1/2}\hbar\omega_0\ell_{\text{HO}})=3.14$.
(c) $g=\infty$. (d) $g/(2^{1/2}\hbar\omega_0\ell_{\text{HO}})=-3.77$.
(e) $g=0-$. (f) The first excited
state of the harmonic oscillator,
$\phi_1(x)$ (blue [dark gray] curve),
being the QPWF of $\left|\uparrow\uparrow
\right>$,
is compared with $\varphi_{\text{QP}}(x)$ of
$\left|\uparrow\downarrow\right>$ at $g=\infty$.
\label{WF}}
\end{figure}

In this Letter a theory of
tunneling for two interacting atoms is developed,
based on the calculation of the \emph{quasiparticle} wave function 
(QPWF) \cite{Rontani05, Toroz11}. The QPWF is the interacting counterpart 
of the single-particle orbital resonating at the energy $\varepsilon$ of
the atom tunneling out of the trap, dressed by the
interaction with the companion atom.
At $g=\infty$, the QPWFs of both $\left|\uparrow\downarrow\right>$ 
and $\left|\uparrow\uparrow\right>$ are found to be identical in the tunnel
barrier [Fig.~\ref{WF}(f)], hence the decay times $\tau$ coincide.
At finite $g$, $\tau$ is evaluated numerically
from the exact wave function for harmonic confinement \cite{Busch}
plus perturbation theory (PT) to account for the anharmonic terms of the
potential trap.   
The two-body correction significantly improves the agreement 
with experimental data [solid lines in Fig.~\ref{tau}(a)],
illustrating a novel few-body effect, similar to
orthogonality catastrophe \cite{Anderson}.

The starting point of the theory is the Hamiltonian 
\begin{equation}
H = -\frac{\hbar^2}{2m}\sum_{i=1}^2
\left[\frac{d^2}{dx_i^2}+V(x_i)\right]+g\delta(x_1-x_2),
\end{equation}
with $m$ being the mass and $V(x)$  
an effective one-dimensional potential. The plot of $V(x)$ in 
Fig.~\ref{frame} reproduces the setup of Ref.~\onlinecite{Zuern11}.
Atom-atom interaction is neglected outside the trap,
that is approximately parabolic at low energy.
Here the two atoms are either spin-1/2 fermions or spinless bosons.
The treatment for bosons is the same as for $\left|\uparrow\downarrow\right>$,
hence only fermions are considered.

Following Bardeen \cite{Bardeen61}, the space is divided into three regions: 
the trap (I), the barrier (II), and the vacuum (III) (Fig.~\ref{frame}). 
The boundaries are the 
classical turning points $x_a$ and $x_b$, hence region II extends from
$x_a$ to $x_b$.
There are two classes of 
single-particle states: 
\newline (i) Vacuum stationary waves 
$\chi_{\varepsilon}(x)$ of (continuous) energy $\varepsilon$.
These are plane waves in region III
reflected at the barrier and 
dropping exponentially with distance into region II.
For example, in the WKB approximation one has 
$
\chi_{\varepsilon}(x)=Ck(x)^{-1/2}\cos{(
\int_{x_b}^x \!\! k(x')dx'-\pi/4 )}$ for $ x>x_b $
and
$\chi_{\varepsilon}(x)
=Ck(x)^{-1/2}2^{-1}\exp{(-\int_x^{x_b}\!\!
k(x')dx')}$ for $ x_a<x<x_b$,
where $C$ is a normalization constant and
$k(x)=[(2m/\hbar^2)\left|\varepsilon-V(x)\right|]^{1/2}$.
For $x<x_a$
$\chi_{\varepsilon}(x)$ drops smoothly to zero, instead of oscillating, 
so $\chi_{\varepsilon}(x)$ is not a good solution for $x<x_a$.
\newline (ii) Trap eigenstates $\phi_n(x)$ ($n=0,1,2,\ldots$),
which have a vanishing tail in region II and smoothly go to zero 
in region III. In region I $\phi_n(x)$ is approximately an eigenstate 
of the harmonic oscillator (HO) with energy
$\hbar\omega_0(n+1/2)$.
\newline 
To summarize, trap states  
are solutions of the single-particle Schr\"odinger equation in regions 
I and II but not in III, whereas stationary waves 
are solutions in regions II and III but not in I. 
Therefore, none of the above states are eigenstates in the whole space,
but they are approximately orthogonal.

Similarly, 
$\Psi_0(x_1,x_2)$ and $\Psi_{m,\varepsilon}(x_1,x_2)$
are states of the entire system with two fermions
that differ in the transfer of an atom from region I to region III.
$\Psi_0(x_1,x_2)$ has two atoms in the trap, either
distinguishable ($\Psi_0=\psi_{\uparrow\downarrow}$)
or indistinguishable ($\Psi_0=\psi_{\uparrow\uparrow}$ 
or $\psi_{\downarrow\downarrow}$).
The latter are non interacting as 
$\psi_{\uparrow\uparrow}(x,x)=0$ for symmetry.
$\Psi_{m,\varepsilon}$ is the non-interacting state with
one atom left in the trap orbital $\phi_m(x)$ and the other one
transferred to the vacuum state $\chi_{\varepsilon}(x)$ (Fig.~\ref{frame}
depicts $\Psi_{0,\varepsilon}$).
Thus $\Psi_0$ is a solution of the interacting 
Schr\"odinger equation with energy $W_0$ in region I and II but not in III,
and $\Psi_{m,\varepsilon}$ is a solution with energy
$W_{m,\varepsilon}$ in region II and III but not in I.
Both $\Psi_0$ and $\Psi_{m,\varepsilon}$ are good solutions in region II.

Bardeen has shown \cite{Bardeen61} that the matrix element 
$M_{m,\varepsilon}$ for the
transition from $\Psi_0$ to $\Psi_{m,\varepsilon}$, as long as the energy 
is conserved ($W_0\approx W_{m,\varepsilon})$, is    
$M_{m,\varepsilon}= -\hbar^2 J_{m,\varepsilon}/2m$,
with $J_{m,\varepsilon}$ being proportional to
the matrix element of
the probability current density operator, 
\begin{equation}
J_{m,\varepsilon}  =  \sum_{i=1}^2
\int_{-\infty}^{\infty}\!\!\!\!\!\!\!\! dx_1 \!\!
\int_{-\infty}^{\infty}\!\!\!\!\!\!\!\! dx_2\,
\,\delta(x_i-x_{\text{bar}})\!
\left[\Psi_0^*\frac{d \Psi_{m,\varepsilon} }{dx_i}-
\Psi_{m,\varepsilon} \frac{d \Psi_0^* }{dx_i}\right],
\label{eq:main}
\end{equation}
and $x_{\text{bar}}$ being any point in the barrier region II.
It is implicit in Eq.~(\ref{eq:main}) 
that $\Psi_0$ and $\Psi_{m,\varepsilon}$
have the same spinorial component as tunneling does not affect
spin. The decay rate $1/\tau$
may be estimated from Fermi golden rule,
\begin{equation}
\frac{1}{\tau}=\frac{2\pi}{\hbar}\sum_{m,\varepsilon}
\left|M_{m,\varepsilon}\right|^2\delta(W_0-W_{m,\varepsilon}).
\label{eq:FGR}
\end{equation}

The non-interacting case $\left|\uparrow\uparrow\right>$
is easy to work out. The trap ground state $\Psi_0$ is the Slater determinant
of the $(n=0)(n=1)$ configuration,
\begin{displaymath}
\psi_{\uparrow\uparrow}(x_1,x_2)=\frac{1}{\sqrt{2}}\left[
\phi_0(x_1)\phi_1(x_2)-\phi_0(x_2)\phi_1(x_1)\right].
\end{displaymath}
Among possible final states, the most relevant one 
is the $(n=0)(\varepsilon=3\hbar\omega/2)$
configuration $\Psi_{0,\varepsilon}$,
\begin{displaymath}
\Psi_{0,\varepsilon}(x_1,x_2)=\frac{1}{\sqrt{2}}\left[
\phi_0(x_1)\chi_{\varepsilon}(x_2)
-\phi_0(x_2)\chi_{\varepsilon}(x_1)\right],
\end{displaymath}
providing the transfer matrix element $J_{0,3\hbar\omega/2}$,
\begin{equation}
J_{0,3\hbar\omega/2}  =  \left[ \phi_1^*(x) 
\frac{d \chi_{\varepsilon}(x) }{dx} -
\chi_{\varepsilon}(x) \frac{d \phi_1^*(x) }{dx}
\right]_{x=x_{\text{bar}}}^{\varepsilon=3\hbar\omega/2}.
\label{eq:Jindist}
\end{equation}
Equation (\ref{eq:Jindist}) is the obvious non-interacting result
for resonant tunneling
between the highest occupied orbital of the trap, $\phi_1(x)$, 
and the vacuum state, $\chi_{\varepsilon=3\hbar\omega/2}(x)$, 
with the other spectator
atom frozen in the orbital $\phi_0(x)$ of the trap. 
In principle, there are final states other than
$\Psi_{0,3\hbar\omega/2}$ allowed by energy conservation,
like the $(n=1)(\varepsilon=\hbar\omega/2)$ configuration.
However, the corresponding matrix elements are negligible
since wave function tails drop exponentially with energy in the barrier.  

The case of two distinguishable
fermions is not trivial, as $\psi_{\uparrow\downarrow}(x_1,x_2)$ 
may not be written
as a single Slater determinant. Nevertheless, 
the exact analytical
solution is available in the relative-motion frame 
\cite{Busch, Girardeau10}. Explicitly, 
\begin{equation}
\psi_{\uparrow\downarrow}(x_1,x_2) =
\Psi_{\text{CM}}([x_1+x_2]/2)\Psi_r(x_1-x_2),
\label{eq:rel}
\end{equation}
with $\Psi_{\text{CM}}$ and $\Psi_r$ being functions of the center-of-mass
and relative-motion coordinates, respectively. 
Since the singlet spin part is an odd function under
particle exchange,
$\Psi_r$ must be even, as shown in Fig.~\ref{WF} (red [gray] curves),
where $\Psi_r(x)$ is plotted for various interaction strengths $g$. 
Note that $\Psi_r(x)$ has two nodes for $g<0$ (super-TG regime).
The wave function 
$\Psi_{\text{CM}}$ is just the lowest
HO state in the center-of-mass frame.

The final state $\Psi_{0,\varepsilon}$ is again a spin singlet, 
with an atom left in the trap lowest orbital
$\phi_0(x)$ and the other one transferred into 
the vacuum state $\chi_{\varepsilon}(x)$, 
\begin{equation}
\Psi_{0,\varepsilon}(x_1,x_2) =
\frac{1}{\sqrt{2}}\left[\phi_0(x_1)\chi_{\varepsilon}(x_2)+
\phi_0(x_2)\chi_{\varepsilon}(x_1)\right].
\label{eq:final}
\end{equation}
Here the resonance energy $\varepsilon$ is fixed by energy conservation.
$\varepsilon$ is plotted in Fig.~\ref{tau}(b) (black curve)
as a function of the magnetic field $B$
that controls the interaction strength $g$ (red [gray] curve) in the 
experiment.

By inserting expressions (\ref{eq:rel}) and (\ref{eq:final}) into
(\ref{eq:main}) one obtains the important result
\begin{equation}
J_{0,\varepsilon}  =  \left[ \varphi_{\text{QP}}^*(x) 
\frac{d \chi_{\varepsilon}(x) }{dx} -
\chi_{\varepsilon}(x) \frac{d \varphi_{\text{QP}}^*(x) }{dx}
\right]_{x=x_{\text{bar}}},
\label{eq:J0QP}
\end{equation}
with $\varphi_{\text{QP}}(x)$ being the QPWF
of the atom escaping from the trap, defined as
\begin{equation}
\varphi_{\text{QP}}(x)=\sqrt{2}\int_{-\infty}^{\infty}\!\!dx'\phi_0(x')
\Psi^*_{\text{CM}}([x'+x]/2)\Psi^*_r(x'-x).
\label{eq:qp}
\end{equation}
From the comparison of Eq.~(\ref{eq:J0QP}) and (\ref{eq:Jindist})
it is clear that,
even in the presence of interaction, the 
tunneling is due to
the resonance between the
vacuum state $\chi_{\varepsilon}(x)$
and the quasiparticle state
$\varphi_{\text{QP}}(x)$. 
With respect to
the non-interacting case of indistinguishable fermions,
$\varphi_{\text{QP}}(x)$
generalizes the role of the highest occupied
orbital $\phi_1(x)$.
Conversely, $\phi_1(x)$ is the QPWF of the trivial 
$\left|\uparrow\uparrow\right>$ case.

At infinite interaction, $g=\infty$, $\Psi_r(x)$ is
`fermionized' [red (gray) curve in
Fig.~\ref{WF}(c)]. This means that $\Psi_r(x)$ overlaps 
with the modulus of the 
wave function of the non-interacting state 
$\left|\uparrow\uparrow\right>$. 
The latter is just $\left|\phi_1\!(x)\right|$ 
in the relative-motion frame. The QPWF 
of $\left|\uparrow\downarrow\right>$ becomes
\begin{eqnarray}
&&\varphi_{\text{QP}}(x) = \frac{2^{1/2}}{\pi^{3/4}\ell_{\text{HO}}^{5/2}}
e^{-x^2/2\ell_{\text{HO}}^2} \quad\times \nonumber\\
&&\left[\int_x^{\infty}\!\!\!\!\!\!dx'
e^{-x'^2/\ell_{\text{HO}}^2}(x'-x)
+\!\!   
\int_{-\infty}^x\!\!\!\!\!\!dx'
e^{-x'^2/\ell_{\text{HO}}^2}(x-x')
\right],
\label{eq:QPgeneric}
\end{eqnarray}
with $\ell_{\text{HO}}
=(\hbar/m\omega_0)^{1/2}$ being the HO length in the laboratory frame.
From Fig.~\ref{WF}(f) it is patent that 
$\varphi_{\text{QP}}(x)$ (black curve)
is unrelated to $\phi_1(x)$ in the laboratory frame
(blue [gray] curve), i.e.,
the QPWF of 
$\left|\uparrow\uparrow\right>$.
Indeed,
$\varphi_{\text{QP}}(x)$ is an even function, with opposite peaks 
separated by a valley with significant weight in the
center, whereas $\phi_1(x)$ has a node.

However, if $x$ is in the potential barrier the first
integral between square brackets in Eq.~(\ref{eq:QPgeneric}) may
be neglected and the second one
integrated over the whole space. This immediately provides
$\varphi_{\text{QP}}(x_{\text{bar}})\cong \phi_1(x_{\text{bar}})$.
As it may be checked visually in Fig.~\ref{WF}(f), this approximate
identity is very well satisfied in region II.
Since the energies $\varepsilon=3\hbar\omega_0/2$ of both QPWFs are the same,
it follows that the transfer matrix elements (\ref{eq:Jindist})
and (\ref{eq:J0QP}) are identical and hence the decay times $\tau$,
in agreement with the experimental data (circles with error bars)
shown in Fig.~\ref{tau}(a). 

The inspection of Figs.~\ref{WF}(a-e) over the whole range of $g$
reveals that $\varphi_{\text{QP}}(x)$ is an amplitude
distibution conceptually different from $\Psi_r(x)$.
This is evident from the very fact that the
two functions have separate natural frames, 
respectively the laboratory 
and the relative-motion frame (note the distinct
lateral sizes in Fig.~\ref{WF}). 
The form of $\varphi_{\text{QP}}(x)$
in the non interacting limits $g=0+$ 
and $g=0-$ may be derived explicitly,
providing respectively $\sqrt{2}\phi_0(x)$ [Fig.~\ref{WF}(a)]
and $\phi_2(x)/\sqrt{2}$ [Fig.~\ref{WF}(e)]. 
Overall, $\varphi_{\text{QP}}(x)$ smoothly
interpolates between these two limits as $1/g$ goes from $+\infty$
to $-\infty$.

The magnitude of $\tau$ depends on
the tail of $\varphi_{\text{QP}}(x)$ in the potential barrier
[Eq.~(\ref{eq:J0QP})].
Since there the QPWF profile   
is similar to those of non-interacting HO orbitals, 
it may seem that interaction affects $\tau$ only by
renormalizing the tunneling energy $\varepsilon$. 
This is incorrect, as
the norm $A_{\text{QP}}$ of the QPWF,
$A_{\text{QP}}=\int dx \left|\varphi_{\text{QP}}(x)\right|^2$,
varies strongly with the interaction strength. 
In fact, $A_{\text{QP}}$ decreases monotonously
from the value 2 at $1/g=+\infty$ down to 1/2 at $1/g=-\infty$,
being one at $1/g=0$.

Therefore, the decay rate $1/\tau$ of Eq.~(\ref{eq:FGR}) 
may be approximately written as
\begin{equation}
\frac{1}{\tau}\cong\frac{A_{\text{QP}}}{\tau_0},
\label{practical}
\end{equation}
with the following provisos: (i) only one final
state is considered, resonating at the energy $\varepsilon$ 
fixed by the interaction strength $g$;
(ii) $\tau_0$ is the decay time obtained in the absence
of interaction 
for elastic tunneling at energy $\varepsilon$; 
(iii) $A_{\text{QP}}$ is the norm of the QPWF whose energy 
is $\varepsilon$.

Figure \ref{tau}(a) shows the single-particle decay time $\tau_0$ 
(dashed lines) as a function of $B$
(interaction strength) 
evaluated through the WKB formula
used in Ref.~\onlinecite{Zuern11}, 
\begin{equation}
\frac{1}{\tau_0}=
\frac{\varepsilon}{2\pi\hbar}\exp{\left(-2\!\!
\int_{x_a}^{x_b}\!\!\!k(x)dx\right)}.
\label{Zuernformula}
\end{equation}
In this approximation   
$\tau_0$ depends only on $\varepsilon$ and the shape of $V(x)$.
For $\left|\uparrow\uparrow\right>$,
$\tau_0$ is independent from $B$ which 
was also observed in the experiment 
[red (gray) circles with error bars in Fig.~\ref{tau}(a)], being
the state non-interacting. For $\left|\uparrow\downarrow\right>$
(black dashed lines and circles) the agreement is only qualitative,
with $\tau_0$ spanning three decades in the 1-1000 msec range 
and decreasing with increasing energy. The departure of
$\tau_0$ from the measured time is systematic, with a blueshift 
on the repulsive side with 
the largest error of $45\%$ at $B=725$ G, and a redshift
on the attractive side, with an error of $65\%$ at 900 G.

The QPWF correction based on
Eq.~(\ref{practical}) improves significantly
the agreement between the predicted [solid black lines 
in Fig.~\ref{tau}(a)] and measured (black circles) values of $\tau$
on the positive side \cite{Lode09,Kim11} of the interaction strength $g$,
as the matching is almost within the error bars.
On the negative side the estimated value
of $\tau$ is strongly blueshifted with respect to $\tau_0$,
with a residual error of 40\% at $B=900$ G.
Such discrepancy originates from the difference
between the actual potential $V(x)$ shown in Fig.~\ref{frame}
and the harmonic trap $V_{\text{HO}}(x)=m\omega_0^2(x-x_0)^2/2$
used to compute $\varepsilon$ \cite{Busch}, 
the higher the energy the stronger the anharmonicity.

By employing PT to first order in $V(x) - V_{\text{HO}}(x)$,
one may evaluate the correction $\Delta \varepsilon$ 
to $\varepsilon$, as shown in Fig.~\ref{tau}(b) (green [light gray] curve). 
Here the free parameter $x_0 =
0.659\ell_{\text{HO}}$ has been fixed by minimizing $\Delta \varepsilon$ 
at $g=\infty$ and the resulting small value 
$\Delta \varepsilon = 0.02\hbar\omega_0$ has been rigidly offset at any 
$g < 0$.
Note that $\Delta \varepsilon \ll \hbar\omega_0$ in the
considered range of $B$,
justifying the usage of PT \emph{a posteriori}. The value of
$\varepsilon$ corrected by PT is then used to compute $\tau_0$ through
Eq.~(\ref{Zuernformula}) [dashed green (light gray) 
line in Fig.~\ref{tau}(a)]
and $\tau$ through (\ref{practical}) [solid green (light gray) line].
Now the predicted value of $\tau$ satisfactorily reproduces 
the experimental data, uncovering the two-body effect inherent 
in the loss of weight of the QPWF.

\begin{figure}
\begin{center}
\includegraphics[trim = 0 30 0 0, width=8.0cm]{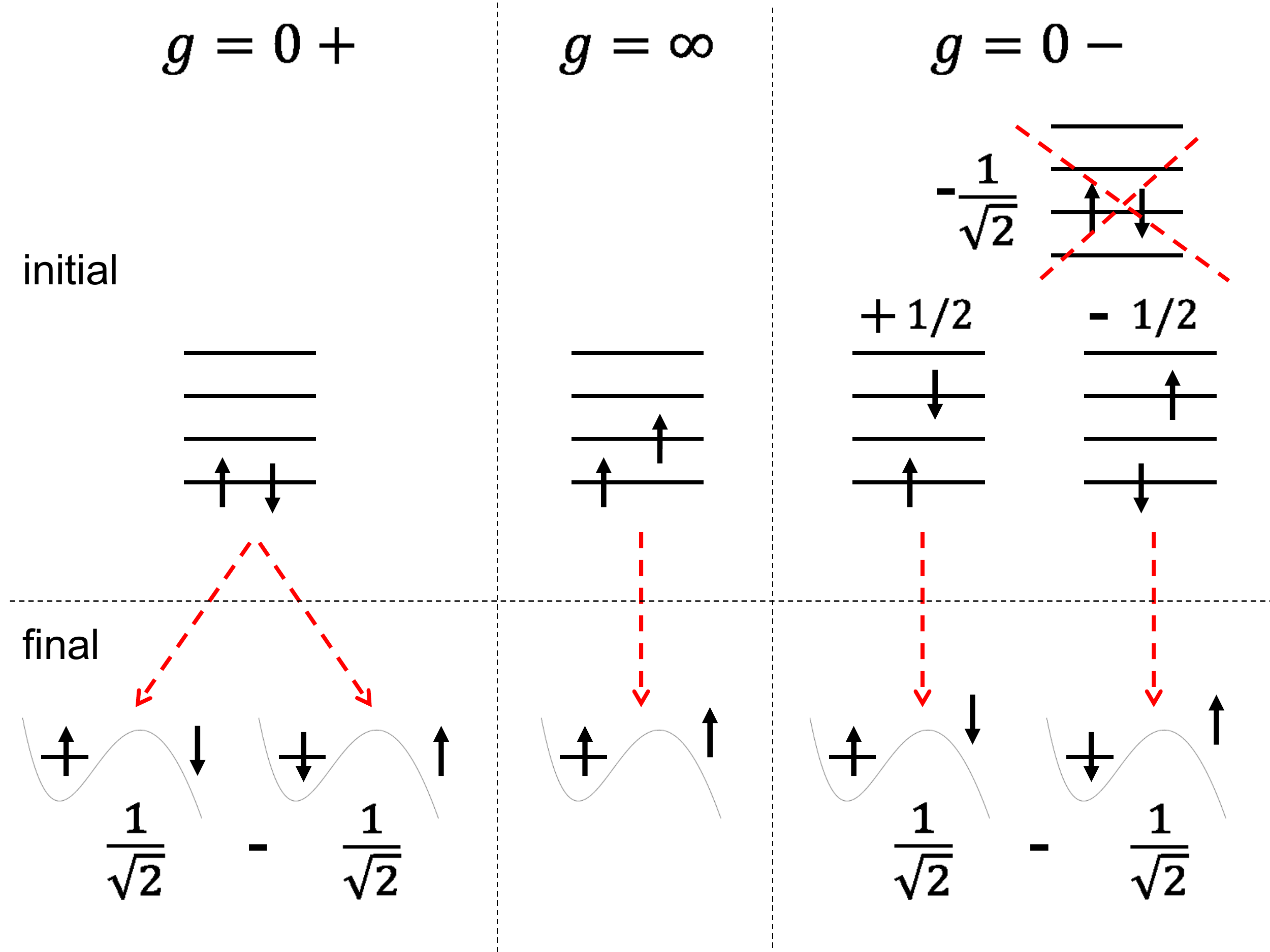}
\end{center}
\caption{(Color online) Expansion of two-fermion states 
at various interaction strengths $g$
on the basis of the Slater determinants obtained by filling with the
atoms the orbitals of the harmonic oscillator (depicted as a ladder)
and the stationary waves in the vacuum. The numbers
are the pertinent expansion coefficients. No label means
a unity coefficient. The dashed red [gray] arrows
point to the allowed tunneling transitions between initial
and final configurations.
\label{configurations} }
\end{figure}

The norm $A_{\text{QP}}$ of the QPWF measures the matching of the initial
two-atom state $\Psi_0$ with that obtained by letting the
current operator act on the final state $\Psi_{0,\varepsilon}$.
To understand the evolution of $A_{\text{QP}}$ with $g$ 
it is useful to write $\Psi_0$ in the laboratory frame.
The simplest case of $\left|\uparrow\uparrow\right>$ is schematized in 
the center column of Fig.~\ref{configurations}.
The initial state is a Slater determinant with one atom in $\phi_0(x)$
and the other one in $\phi_1(x)$; 
the final state 
has one atom in $\phi_0(x)$ and one in $\chi_{\varepsilon}(x)$.
The coefficients of both
determinants are one. The two states overlap if one transfers
one atom from $\phi_1(x)$ into $\chi_{\varepsilon}(x)$, as indicated
by the dashed arrow.
This single-particle result 
may be taken as a reference, providing
a unity contribution to the tunneling matrix element.
This discussion holds also for 
$\left|\uparrow\downarrow\right>$ at $g=\infty$.

The continuous evolution of $A_{\text{QP}}$ from $g=-\infty$
to $g=0-$ may be understood by considering the
limit of $g=0-$, depicted in the right column 
of Fig.~\ref{configurations} (not accessible in the experiment). 
Three different configurations
span $\Psi_0$ in the lab frame: the two atoms
either doubly occupy $\phi_1(x)$ 
or singly occupy $\phi_0(x)$ and $\phi_2(x)$ with alternate spins. 
However, only
the latter configurations are effective in transferring the
atom from $\phi_2(x)$ into $\chi_{\varepsilon}(x)$, 
providing two non-zero spin-conserving contributions, 
each of weight $1/2\times 1/\sqrt{2}$. 
The total contribution to the squared matrix element is 1/2.
Such decrease of the QPWF weight, arising from the multiple configurations
that generically span a many-body state, is familiar from the
field of strongly correlated electrons interacting 
through Coulomb repulsive forces \cite{Rontani05}.

The positive branch of $g$ shows a more exotic behavior. 
In the limit $g=0+$ (left column of Fig.~\ref{configurations})
$\Psi_0$ is a non-interacting configuration with $\phi_0(x)$
being doubly occupied. This single configuration
contributes twice to the final state, as both atoms may tunnel
with opposite spins ($2\times 1/\sqrt{2}$).
Therefore, the square matrix element is \emph{doubled}.
This coefficient may be regarded as a bosonic-like enhancement factor.
Indeed, $\psi_{\uparrow\downarrow}(x_1,x_2)$ is the wave 
function of two spinless non-interacting bosons, both occupying the
level $\phi_0(x)$.

In conclusion, the quasiparticle theory of tunneling 
here developed 
explains the available experimental data for two  
$^6$Li atoms. The comparison with measured
decay times discloses a genuine two-body effect, related
to the actual form of the interacting wave function.
Therefore, such measurements may be regarded as
a novel type of spectroscopy for cold atoms, paralleling
successful techniques of condensed matter, such as photoemission
and single-electron tunneling. 
 
\begin{acknowledgments}
I thank Gerhard Z\"urn, Friedhelm Serwane, Selim Jochim
for exciting discussions and for making available
their experimental data prior to publication.
I gratefully acknowledge support by Fondazione Cassa di 
Risparmio di Modena through the
project COLDandFEW and from CINECA through 
CINECA-ISCRA project no. HP10BIFGH8. 
\end{acknowledgments}


\end{document}